\documentclass[12pt]{article}

\usepackage{epsfig}
\usepackage{cancel}

\usepackage{centernot}
\usepackage{comment}
\usepackage{setspace}
\usepackage{color}

\def\be{\mbox{\bf e}}

\def\bt{\mbox{\bf t}}

\def\bphi{\mbox{\boldmath $\phi$}}

\def\balpha{\mbox{\boldmath $\alpha$}}

\usepackage{amssymb}

\begin{document}

\begin{titlepage}

\baselineskip 24pt

\begin{center}

{\Large {\bf Search for new physics in semileptonic decays of $K$ 
and $B$ as implied by the $g - 2$ anomaly in FSM}}

\vspace{.5cm}

\baselineskip 14pt

{\large Jos\'e BORDES \footnote{Work supported in part by Spanish MINECO under grant PID2020-113334GB-I00/AEI/10.13039 /501100011033
and CIPROM/2022/36 (Generalitat Valenciana).}}\\
jose.m.bordes\,@\,uv.es \\
{\it Departament Fisica Teorica and IFIC, Centro Mixto CSIC, Universitat de 
Valencia, Calle Dr. Moliner 50, E-46100 Burjassot (Valencia), 
Spain}\\
\vspace{.2cm}
{\large CHAN Hong-Mo}\\
hong-mo.chan\,@\,stfc.ac.uk \\
{\it Rutherford Appleton Laboratory,\\
  Chilton, Didcot, Oxon, OX11 0QX, United Kingdom}\\
\vspace{.2cm}
{\large TSOU Sheung Tsun}\\
tsou\,@\,maths.ox.ac.uk\\
{\it Mathematical Institute, University of Oxford,\\
Radcliffe Observatory Quarter, Woodstock Road, \\
Oxford, OX2 6GG, United Kingdom}

\end{center}

\vspace{.3cm}

\begin{abstract}
The framed standard model (FSM), constructed to explain, with some 
success, why there should be 3 and apparently only 3 generations of 
quarks and leptons in nature falling into a hierarchical mass and 
mixing pattern \cite{tfsm}, suggests also, among other things, a 
scalar boson $U$, with mass around 17 MeV and small couplings to 
quarks and leptons \cite{cfsm}, which might explain \cite{fsmanom} 
the $g - 2$ anomaly reported in experiment \cite{gminus23}.  The $U$ 
arises in FSM initially as a state in the predicted ``hidden sector"
with mass around 17 MeV, which mixes with the standard model (SM) 
Higgs $h_W$, acquiring thereby a coupling to quarks and leptons and 
a mass just below 17 MeV.  The initial purpose of the present paper 
is to check whether this proposal is compatible with experiment on 
semileptonic decays of $K$s and $B$s where the $U$ can also appear.  
The answer to this we find is affirmative, in that the contribution 
of $U$ to new physics as calculated in the FSM remains within the 
experimental bounds, but only if $m_U$ lies within a narrow range 
just below the unmixed mass.  As  a result from this, one has an 
estimate $m_U \sim 15 - 17$ MeV for the mass of $U$, and from some 
further considerations the estimate $\Gamma_U \sim 0.02$ eV for its 
width, both of which may be useful for an eventual search for it in 
experiment.  And, if found, it will be, for the FSM, not just the 
discovery of a predicted new particle, but the opening of a window 
into a whole ``hidden sector" containing at least some, perhaps even the bulk,
 of the dark matter in the universe.

\end{abstract}  

\end{titlepage}

\newpage

\section{Preamble}

Deviations from the standard model, known respectively as the 
$g - 2$ and Lamb shift anomalies, have long been reported in 
experiments on the magnetic moment of the muon \cite{gminus21,gminus22}
and on the Lamb shifts in muonic hydrogen \cite{LsanomH1,LsanomH2} and 
deuterium \cite{LsanomD1,LsanomD2}, and many authors \cite{lowmscalarsa,lowmscalarsb} 
have already noted that they could be explained by a scalar 
boson of low mass with small couplings to leptons and quarks.  
We ourselves have made the observation \cite{fsmanom} that a 
scalar particle called $U$, which is predicted by the framed 
standard model (FSM) \cite{tfsm} we work on, would also serve 
this purpose.   

A distinguishing feature of our suggestion is that the scalar 
boson $U$ was not created just for the purpose of explaining 
the $g - 2$ and Lamb shift anomalies but was predicted by the 
FSM constructed to try and resolve the old generation puzzle why 
leptons and quarks should exist in 3 (and apparently only 3) 
generations, and why they should fall in such a peculiar mass 
and mixing pattern.  The $U$ originates there as a particular 
scalar state, called $H_+$ in \cite{cfsm} with mass $\sim$ 17 
MeV predicted by the FSM in the hidden (dark) sector which mixes 
with the standard model (SM) Higgs $h_W$.  The mixed state $U$ 
is expected thus to have a mass close to 17 MeV and a small 
component in $h_W$, acquiring through which a coupling to quarks 
and leptons, as wanted to explain the anomalies.  How this comes 
about is of necessity a longish story to which we shall provide 
a brief outline in the next section sufficient for the present
paper.  For more details, however, the reader will have to be 
referred elsewhere. 

Notice that the FSM predicts only the existence of this mixed 
state $U$ but is at present unable, for reasons to be detailed 
later, to actually solve the mixing problem to give the values 
of the mixing parameter $\rho_{Uh}$ and the mass $m_U$ of the 
mixed state $U$.  It is shown in \cite{fsmanom}, however, that 
by choosing the values of $\rho_{Uh}$ and $m_U$ judiciously, 
one can reproduce the experimental results for both anomalies.  
This result serves thus a dual purpose: 
\begin{itemize}
\item {\bf [a]} 
It uses the FSM to explain the $g - 2$ and Lamb shift anomalies 
reported in experiment. 
\item {\bf [b]} 
It uses the experimental data on the anomalies to put bounds on 
the parameters $\rho_{Uh}$ and $m_U$ of $U$ which the FSM is at 
present unable to provide. 
\end{itemize}
Henceforth, we shall refer to this scenario as the ``$U$ proposal" 
of the FSM for short. 

Since then, however, although the $g - 2$ anomaly is confirmed 
by new experiment \cite{gminus23} the Lamb shift anomaly is not  
\cite{bezginov,xiong}, so that the analysis for the latter has to be 
redone when the data settle.  Until that is done, therefore, as 
concerns the FSM, this means that part of {\bf [b]} is put in 
abeyance, in that the $g - 2$ anomaly puts constraint only on 
$\rho_{Uh}$, hardly on $m_U$, so that there is no check yet from 
data on the $U$ mass.  Hence, in what follows, we shall refer to 
\cite{fsmanom} only for the result on $g- 2$, not for that on 
the Lamb shift.  

Whatever will happen to the Lamb shift anomaly, however, so long 
as the $U$ is proposed for the $g - 2$ anomaly, the story should 
not stop here, for the $U$ can occur also in other reactions, and 
it would be incumbent upon us to check in the FSM whether the effects 
resulting from $U$ are consistent with the existing data in these 
other reactions.  This is what concerns us in the present paper.

\begin{figure}[t]
\centering
\includegraphics[scale=0.6]{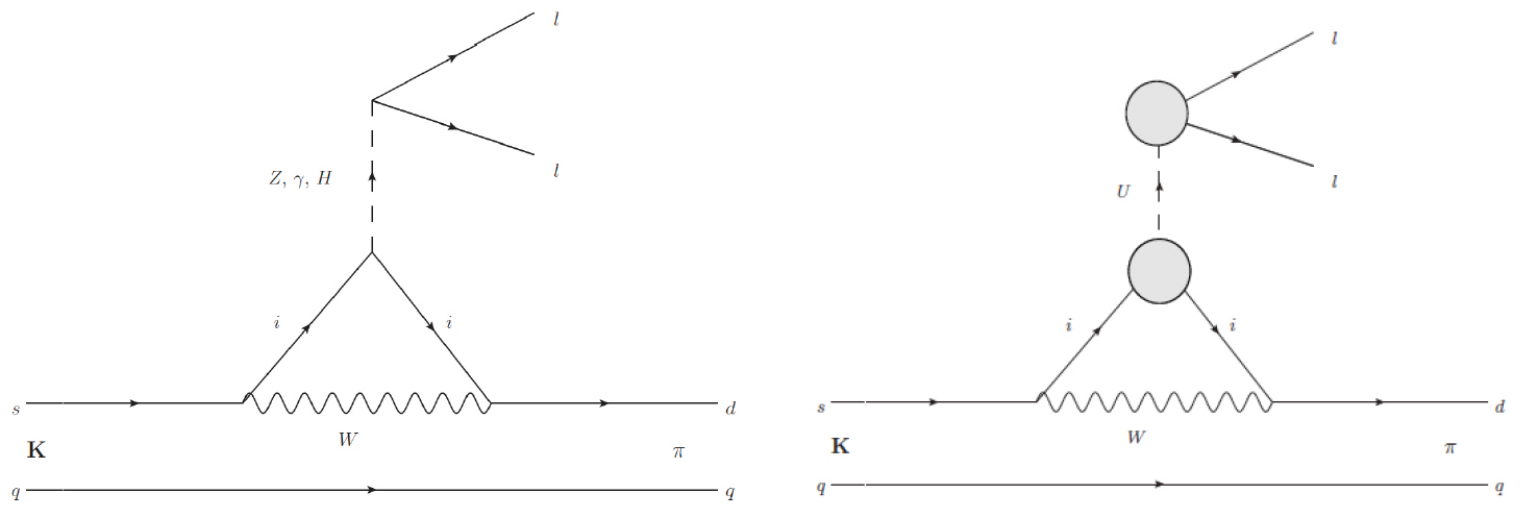}
\caption{Left panel: SM penguin contribution to the decay of kaons, $K \rightarrow \pi \ell \ell$ in the spectator model. Right panel: additional contribution from $U$ in FSM.}
\label{penguin-comb}
\end{figure}

We focus first on the semileptonic decays of the $K$ and $B$ mesons 
which are a favourite venue to look for new physics because the SM 
contributions there are strongly suppressed. For instance the `penguin' diagram as that 
shown in Figure \ref{penguin-comb} (left) is first affected by 
the loop, and further by the CKM matrix elements linking the various 
quark states.  This may thus, it is thought, make it easier for any 
new physics, such as the contribution from the $U$ boson suggested 
above, to reveal itself and be identified as such.

The $U$ boson can contribute to semileptonic decays of $K$s and $B$s 
through essentially the same diagram Figure \ref{penguin-comb} (right)
as the SM, but differing in that the $U$ occurs in the physical region 
of, say, $K(B) \rightarrow  \pi(K) e^+ e^-$ so that any significant 
contribution from $U$ will appear as a sharp peak at $m_U$ in the 
invariant mass plot of $e^+ e^-$.  No such peak has ever yet 
been reported.  Even if, for some technical reasons, such a peak is 
hard to detect, the $U$ will contribute to the semileptonic decay 
rate over and above that from the SM and show up as new physics.  
However, the data at present are broadly in agreement with SM 
expectations, showing some possible deviations but yet within the 
experimental bounds.  It would thus be a very stringent test for 
the $U$ proposal in the FSM since its contribution there is in 
principle calculable with no freedom, once given the value of $m_U$ 
and $\rho_{Uh}$.  This is especially the case since $\rho_{Uh}$, the 
mixing parameter, as estimated from the $g - 2$ anomaly, is of order 
0.1 \cite{fsmanom}, quite a bit larger than what most other scalar 
bosons models suggest \cite{lowmscalarsa,lowmscalarsb}. 

Fortunately, as we shall show in what follows, the $U$ proposal does 
manage to pass this test and remain within the present experimental 
bounds, but, quite intriguingly, only for the mass $m_U$ of $U$ taking 
values within a narrow range of merely a few MeVs below the predicted 
unmixed value of 17 MeV, that is, just about where it is expected to be.  
The reason is that the loop in the diagram Figure \ref{penguin-comb} (right) 
is kinematically dominated by the $t$ quark, and it so happens that 
the coupling of $t$ to $U$, as obtained from the FSM, is close to 
zero for some value of $m_U$ close to 17 MeV.  

Now, if it is not just fortuitous but is the real reason why the $U$ 
is not seen in semileptonic decays of $K$s and $B$s, then the FSM 
would seem to have a higher degree of self-coordination, and to a 
higher accuracy, than we suspect or have bargained for initially in 
its construction. 

In other words, one finds that, like the earlier \cite{fsmanom}, the 
present analysis serves also a dual purpose:
\begin{itemize}
\item {\bf [a$'$]} It shows that the $U$ proposal of the FSM used 
before to explain the $g - 2$ anomaly is compatible with present 
data on semileptonic decays of $K$s and $B$s.
\item {\bf [b$'$]} It shows that the present data on these decays 
provide a constraint on the $U$ mass to within a few MeV of the 
unmixed value 17 MeV predicted.
\end{itemize}
This last is much sharper than {\bf [b]} obtained in \cite{fsmanom} 
from the Lamb shift anomaly before.

Further, it will be shown that if one feeds in the data on the mode 
$K^+ \rightarrow \pi^+ +$ missing energy, the ``golden channel", one 
obtains in addition the following: 
\begin{itemize}
\item {\bf [c$'$]}
An estimate of the $U$ width, and a possible lower bound in FSM of 
the lightest dark matter particle,
\end{itemize}
both of which could be of wider interest and potential significance. 

In the following sections, we first outline how $U$ emerged from 
the FSM, then give details of our analysis on semileptonic decays 
of $K$s and $B$s, but end with a critique on what could be a weak 
assumption in both, which, though generally accepted elsewhere and 
when extended to FSM seems practically so far to have worked, may 
need, we think, a closer theoretical scrutiny.

\section{The FSM and $U$}

We shall not try to summarize the FSM in the limited space available 
here, but merely to give an outline with some essentials needed for 
the present paper.  A fairly recent review of the FSM can be found 
in \cite{varyym} apart from 2 new results of substance reported in 
\cite{Higgcoup} and \cite{cevt}.

The FSM was constructed with the aim of explaining why there are in 
nature 3 and only 3 generations of quarks and leptons, and why they 
should fall into such a hierarchical mass and mixing pattern.  In 
this, the FSM has succeeded quite well.  By taking generation as 
dual colour (hence 3 and only 3), it has provided us with not only 
a qualitative understanding why the experimentally observed mass 
and mixing pattern should occur, but even a good fit to the known 
mass and mixing parameters \cite{tfsm}, often to within existing 
errors, with just 7 parameters, thus effectively replacing by these 
few some 17 of the SM's parameters taken from experiment.  Besides, 
the FSM has produced a growing list of unsolicited bonuses not 
envisaged when it was constructed: {\bf [A]} a new solution without 
axions of the strong CP problem together with an answer to why a 
CP-violating phase should appear in the mixing matrix of quarks 
\cite{atof2cps} and leptons \cite{cpslept,cpslash}; {\bf [B]} that 
though sharing the SM's results \cite{Higgcoup} in areas where the 
SM has been applied with success, it can also accommodate those 
deviations from the SM recently reported in experiment such as the 
$g - 2$ anomaly already mentioned; and lately even {\bf [C]} a 
possible new take on the horizon problem in cosmology \cite{cevt}.
 
The FSM is obtained by framing the SM, that is, by promoting frame 
vectors in the internal symmetry $G = u(1) \times su(2) \times su(3)$ 
of the SM to dynamical field variables (framons), similar in spirit 
to vierbeins in gravity.  Frame vectors can be thought of as columns 
of the transformation matrix relating the local (space-time point 
$x$-dependent) frame to a global ($x$-independent) reference frame, 
and carry thus, as vierbeins do, both a local and a global index. 
In other words, they form representations of $G \times \widetilde{G}$ 
where $\widetilde{G}$, which we shall call the dual of $G$, represents 
transformations on the global reference frames.  And since physics 
should be invariant under changes both in the local and in the global 
reference frames, the theory with framons will have to be invariant 
under $G \times \widetilde{G}$. 

There is some freedom in choosing this double representation and the 
FSM has chosen the representation ${\bf 1} \times ({\bf 2} + {\bf 3})$ 
for $G$ but the representation ${\bf \tilde{1}} \times {\bf \tilde{2}} 
\times {\bf \tilde{3}}$ for $\widetilde{G}$.  This choice seems to be the 
only one that works, and it happens also to be a choice which requires 
the smallest number of new scalar fields to be introduced, meaning 
that the FSM is in a sense a minimally framed version of the SM.  In 
any case, with this choice of representation, the framon of FSM breaks 
into two, a `flavour framon' and a `colour framon'. 

The flavour framon takes the form:
\begin{equation}
\balpha \bphi(x)
\label{FF}
\end{equation} 
where $\bphi(x)$ is a flavour $su(2)$ doublet field, and $\balpha$ an 
$x$-independent dual colour $\widetilde{su}(3)$ triplet, which in all 
applications so far of the FSM can be taken real, namely as a unit 
3-vector in what is to be become the generation space for leptons and 
quarks.  This means that the flavour framon (\ref{FF}) is basically 
the same as the Higgs field in the SM but carrying in $\balpha$ 
already a generation index.  It leads then to a Yukawa coupling of 
the form \cite{tfsm}, general to all quarks and leptons:
\begin{equation}
\rho_T | \balpha \rangle (\zeta_W +  h_W) \langle \balpha | ,
\label{Yukawaw}
\end{equation}
and a fermion mass matrix of the form:
\begin{equation}
m = m_T | \balpha \rangle \langle \balpha |,
\label{mfact}
\end{equation}
where $\balpha$, coming from the framon, is common to all quarks and 
leptons, $\zeta_W$ is the vacuum expectation value
(vev) of the Higgs field, $h_W$ is the Higgs 
boson, while $\rho_T$ is the coupling and $m_T = \rho_T \zeta_W$ the 
mass parameter peculiar to the fermion type $T$, that is, whether 
up-type or down-type quarks or charged leptons or neutrinos.

The colour framon, on the other hand, is of a more complicated form, 
which is not exhibited because not needed here, and represents many 
new degrees of freedom with no analogue in the SM.  Previous analyses 
\cite{cfsm} have shown that it will give a new `hidden sector' of 
particles over and above those in the `standard sector' we already 
know: namely the quarks and leptons and the gauge bosons and Higgs.  
This new sector communicates but little with the standard sector we 
live in, hence the label `hidden', and comprises some members which 
can serve as candidate constituents of dark matter.  But there are 
two channels known through which the hidden sector can affect our 
world.

\begin{figure}[th]
\centering
\includegraphics[scale=0.45]{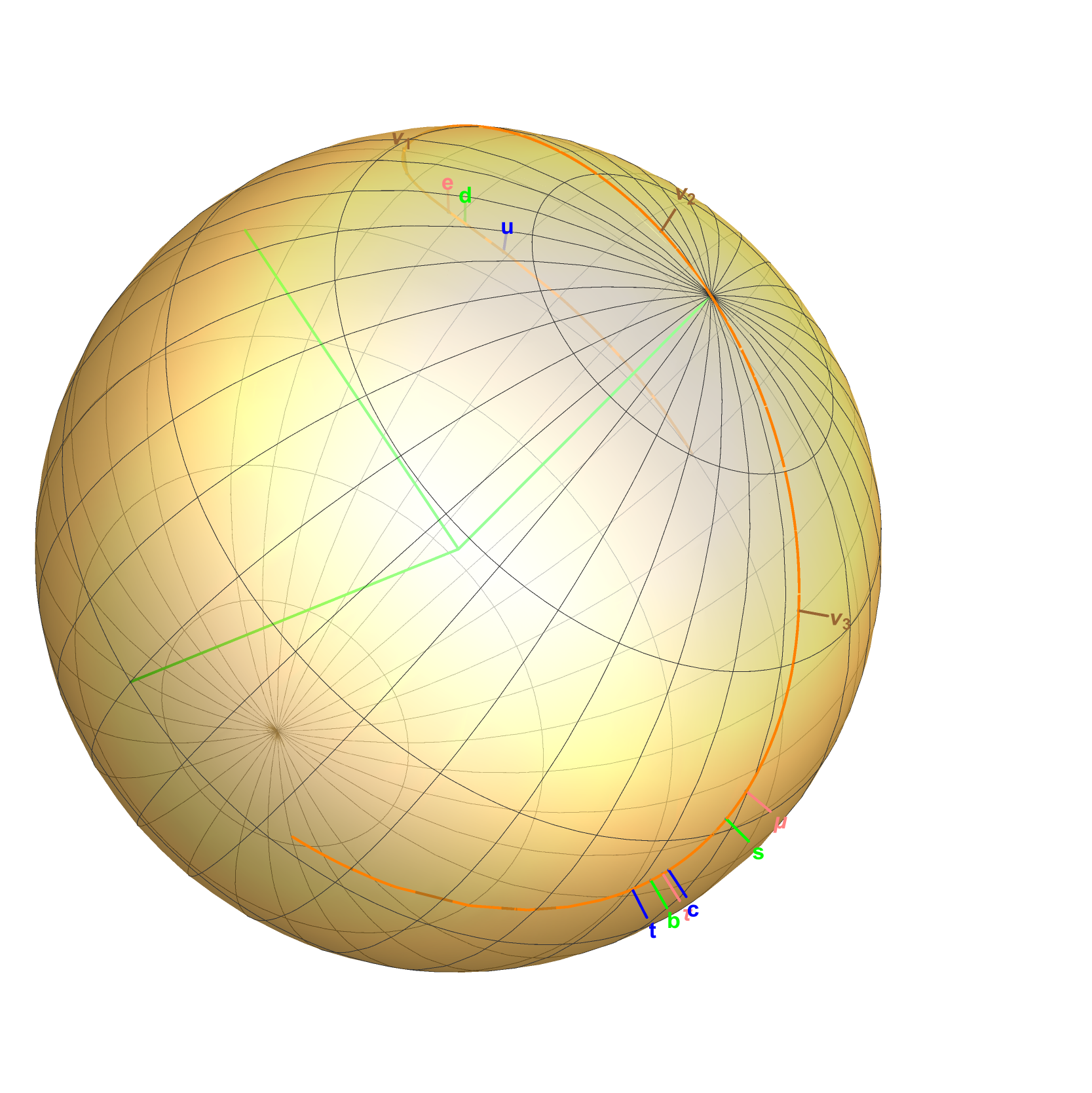}
\caption{The fitted trajectory for $\balpha$ on the unit sphere}
\label{florosphere}
\end{figure}
\begin{itemize}

\item{\bf [I]}
The colour framons being a double representation both of colour and 
dual colour, renormalization by colour framon loops can change the 
relative orientation of the vacuum in colour and dual colour.  Hence 
the vector $\balpha$ appearing in (\ref{mfact}) above which is attached
to the vacuum will be dragged along and rotate with change of scale.  
And in the FSM, it is this rotation of $\balpha$ which gives rise to 
the hierarchical mass and mixing pattern of quarks and leptons seen 
in experiment.  Indeed, by adjusting the few parameters left free by 
the RGEs, a very good fit is obtained \cite{tfsm} for the existing 
data resulting in the rotation trajectory for $\balpha$.  The curve 
traced out by $\balpha$ on the unit sphere as given by the equation:
\begin{equation}
\cos \theta \tan \phi = a
\label{curvetraced}
\end{equation}
with the fitted value $a = - 0.1$ is shown in Figure \ref{florosphere} 
and the speed in $t = 2 \ln \left(\frac{\mu}{1 {\rm Gev}} \right)$ with which it is traced is shown in 
Figure \ref{thetamu}.  To both of these figures we shall later refer.

\begin{figure}[th]
\centering
\includegraphics[scale=0.4]{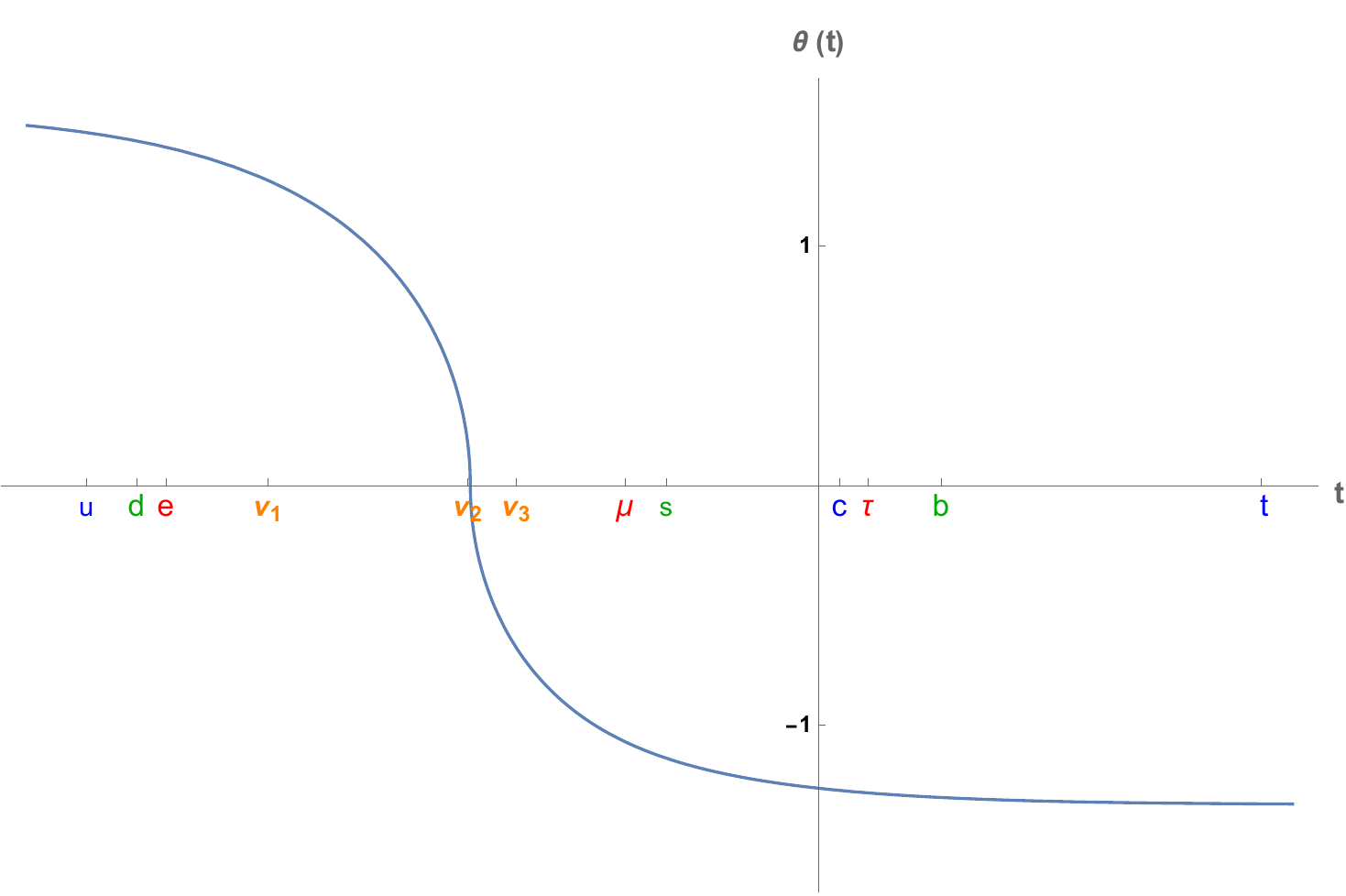}
\caption{Fitted $\theta$ as a function of $t = 2 \ln \left(\frac{\mu}{1 {\rm Gev}} \right)$}
\label{thetamu}
\end{figure}

\item{\bf [II]}
Some of the hidden sector particles are found to mix with particles 
in the standard sector \cite{cfsm}, acquiring thereby small couplings 
therein, hence leading to deviations from the SM.  Of these, the $U$ 
under consideration in this paper, which supposedly leads to the 
$g - 2$ anomaly, is one example.  It has been shown that some other 
possible departures from the SM recently suggested in experiment, 
such as (i) the Lamb shift anomaly (as then known) \cite{LsanomH1,LsanomH2,LsanomD1,LsanomD2}, 
(ii) the Atomki anomaly \cite{Atomki1,Atomki2}, and (iii) a mass for the $W$ 
greater than expected in the SM \cite{mW1,mW2}, can also be accommodated 
in the FSM in this way \cite{fsmanom,zmixed}.

\end{itemize}

Specifically, the $U$ emerges in the FSM as follows.  There is in 
the hidden sector a number of scalar bosons generically labelled 
as $H$ \cite{cfsm}, which are the analogues of the standard Higgs 
boson $h_W$ in the standard sector, only with the roles of colour 
and flavour interchanged.   Their mass matrix is explicitly given in 
\cite{cfsm}, as deduced from the framon self-interaction potential
which, we recall, has to be invariant under $G \times \widetilde{G}$ and 
therefore restricted to a particular form.  The analysis done there 
found that a bunch of these $H$ states have a mass of around 17 MeV,
where the FSM vacuum undergoes a vacuum transition called VTR1, as 
is detailed in \cite{cevt}.  And among this bunch, one, called $H_+$ 
in \cite{cfsm}, is distinguished by being linked to the standard 
Higgs state $h_W$ by an off-diagonal element, and hence will mix 
with it, giving the lower of the two mixed states as the $U$ we 
seek.

Unfortunately, the known information does not allow one as yet to 
solve the mixing problem, depending as it does on some parameters 
which are at present unknown and which besides are scale-dependent 
and it is not clear at which scale (or scales) they are to be 
evaluated.  One has thus, as we do here, to treat both the mixing 
and the $U$ mass as parameters to be determined phenomenologically 
from data, so long as the latter is constrained to remain in the 
vicinity of the unmixed value of 17 MeV.

\begin{figure}[th]
\centering
\includegraphics[scale=0.55]{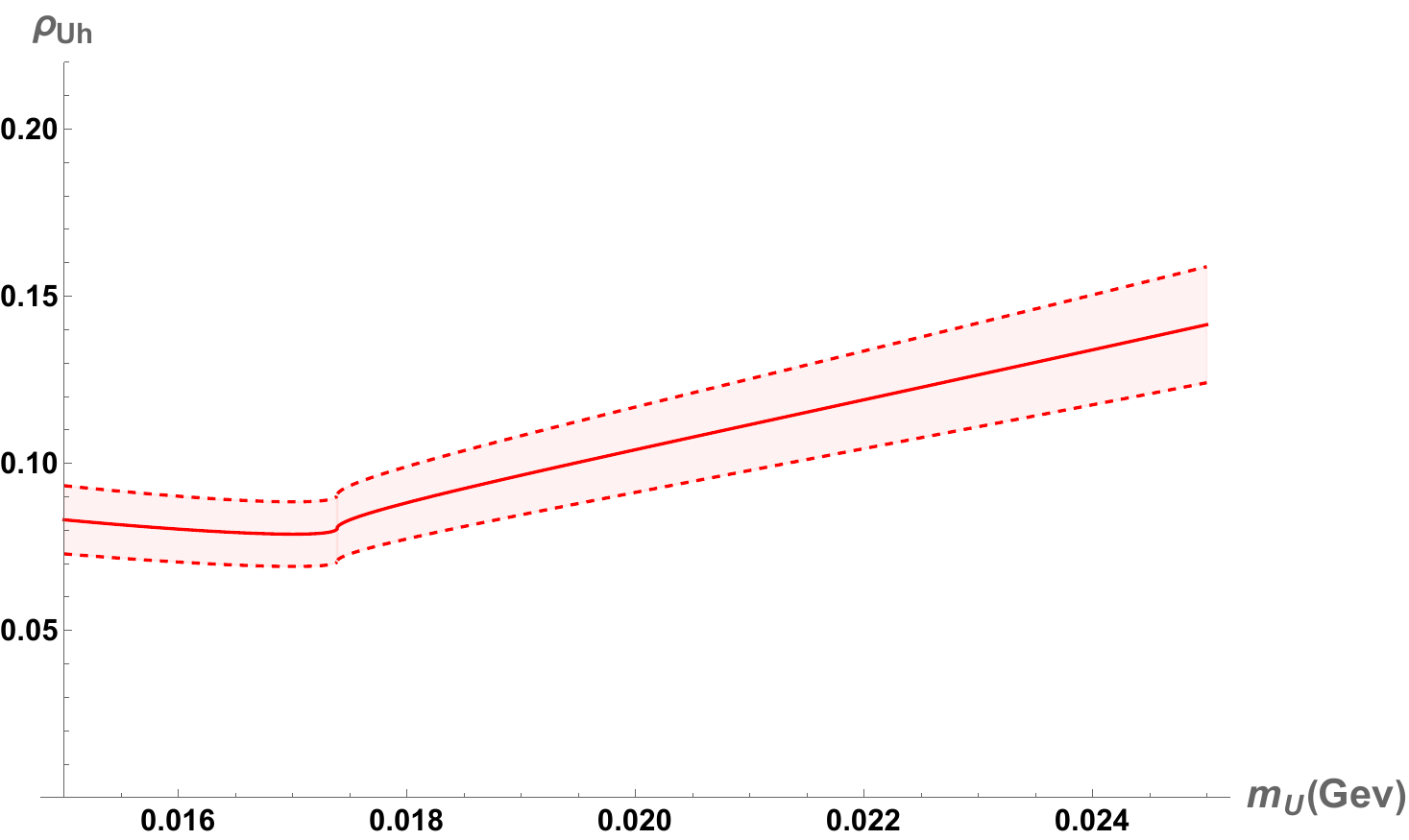}
\caption{Central value (solid line) and allowed experimental bounds on the mixing parameters  $\rho_{Uh}$ 
from muon $g-2$ \cite{fsmanom}.}
\label{rhoUh}
\end{figure}

We recall that in \cite{fsmanom}, we have from the data on $g - 2$ 
Figure \ref{rhoUh}, where the mixing parameter $\rho_{Uh}$ 
is constrained to have a value $\sim 0.1$ but the $U$ mass $m_U$ is 
hardly constrained.  It was the overlap of Figure \ref{rhoUh} with 
the constraint from the Lamb shift data which gave the estimate that 
$m_U \sim 20$ Mev, but this has at this juncture to be ignored until 
the new data are assimilated.  We shall see later what the data on 
semileptonic decays of $K$s and $B$s have to say on $m_U$.

\begin{table}[th]
\begin{spacing}{1.5}
\begin{eqnarray*}
\begin{array}{||c|c r | c r ||}  
\hline \hline\\

\mathrm{Process} & Br \, \mathrm{SM  \, result} & & \mathrm{Experimental}\,Br  &   \\  
\hline \hline
 K^+ \rightarrow  \pi^+ \, \cancel{E} \, &  \left( 8.6 \pm 0.42  \right) \times 10^{-11} & \cite{Buras22} & \left( 1.14  \begin{array}{c}+ 0.40 \\ - 0.30 \end{array} \right)
  \times 10^{-10} & \cite{NA62} \\ 
\hline 
 K_L \rightarrow \pi^0 \, \cancel{E}  \,& \left( 2.94  \pm 0.15  \right) \times 10^{-11} & \cite{Buras22} &  < 3 \times 10^{-9} 
 &  \cite{KOTO} \\ 
\hline
 K_L \rightarrow \pi^0 \,e^+ \, e^-  \, & \left(  3.54 \begin{array}{c}+ 0.98 \\ - 0.85 \end{array} \right) \times 10^{-11} &\cite{Buras2203}  & < 5.1 \times 10^{-10}  & \cite{KTev,nir} \\ 
\hline
B^+ \rightarrow K^+ \, \cancel{E} & \left( 4.4 \pm 0.7 \right)\times 10^{-6} & \cite{HeValencia} &  < 1.6 \times 10^{-5} & 
\cite{lees} \\ 
\hline
B \rightarrow K^{\star} \, \cancel{E} & \left( 9.5 \pm 1 \right) \times 10^{-6} & \cite{HeValencia} &  < 4  \times 10^{-5}  &
\cite{lutz} \\ 
\hline
 B^+ \rightarrow K^+ \,e^+ \, e^- &  \left( 3.6 \pm 1.2 \right) \times 10^{-7} & \cite{074020}  &  \left( 5.6 \pm 0.6 \right) \times 10^{-7} & \cite{aubert,choudhury} \\ 
 \hline
 B \rightarrow K^{\star} \,e^+ \, e^- &  > \, \left( 9.6 \pm 1  \right) \times 10^{-7} & & \left(  1.55 \begin{array}{c}+ 0.40 \\ - 0.31 \end{array} \right) \times 10^{-6} & \cite{aubert,wei} \\ 
 B \rightarrow K^{\star} \,\mu^+ \, \mu^- &  & & \left(  9.6 \pm 1 \right) \times 10^{-7}  & \cite{PDG}\\ 
 \hline \hline
\end{array}
\end{eqnarray*} 
\caption{Branching ratios of the semileptonic decays of $K$ and $B$ mesons for the full phase space region.}
\label{experimentaltable}
\end{spacing}
\end{table}

\section{Semileptonic decays of $K$s and $B$s in SM}

As noted already in the Preamble, semileptonic decays of $K$s and $B$s 
are a favourite venue to look for departures from the SM or new physics.  
These decays are thought to proceed mainly via the penguin diagram shown 
on the left panel of Figure \ref{penguin-comb} which is suppressed first 
by the loop and secondly by the occurrence twice of CKM mixing elements, 
hence affording, it is thought, a good chance for any new physics to 
reveal itself.

The current situation is summarised in Table \ref{experimentaltable}, 
where we note the following:
\begin{itemize}
\item Data for the $K^+$ and $K_S$ semileptonic decays into electrons 
are not included since full theoretical information in the SM 
is not available.
\item When the final state is generically written as ``missing energy 
($\cancel{E}$)" the only final state in the SM are neutrino pairs.  
But in models including new physics (such as the contributions from 
the $U$ boson considered in this paper) there can be other neutral 
particles undetected, like dark matter particles such as `co-neutrinos' 
in the FSM \cite{cfsm}.
\item The data for the decay $B \rightarrow K \mu^+ \mu^-$ are not 
included, not being relevant here since the $U$ of mass $\sim$ 17 MeV 
is insufficiently massive to decay into $\mu^+ \mu^-$.  But the data 
for $B \rightarrow K^* \mu^+ \mu^-$ is included for a special purpose,
namely to serve as a stand-in for the SM result for $B \rightarrow 
K^* e^+ e^-$, of which we found no direct estimate in the literature.
The SM predicts lepton flavour universality in semileptonic decays 
(up to small corrections of order $\left(\frac{m_\mu}{m_b}\right)^2$). 
Hence, above the muon production threshold the branching ratios for 
electrons and muons would be the same.  This seems to be experimentally 
confirmed in recent LHCb results for energies above the muon production 
threshold \cite{newresults}.  (One notes in passing that this reverses 
an earlier result \cite{oldresulta,oldresultb} suggesting violations
of lepton universality.)  In this context then we feel free to take 
the experimental data for muons as a lower bound of the SM result for 
electrons, assuming that discrepancy between the two occurs only at 
invariant mass of the lepton pair below the muon threshold. 
\end{itemize}

One sees that the SM gives a pretty good explanation for the decay 
processes listed, although the theoretical values are plagued with 
uncertainties coming from the hadron structure.  In some channels, 
experiment gives a somewhat larger value than that predicated by 
the SM, but still within present errors, while in others there are 
but bounds.  As a result there is still some room for new physics, 
but not much.

\section{The $U$ contribution to these decays}

The penguin diagram of Figure \ref{penguin-comb} (left), as the main 
source for semileptonic decays of $K$s, has been studied by many 
authors.  Among the pioneers were Leutwyler and Shifman who gave the 
answer for the process $K \rightarrow \pi + \mathrm{boson}$, say $X$, 
in terms of an SM effective Lagrangian \cite{LeutwylerShifman}:
\begin{equation}
\mathcal{L}_{sdU}^{SM} \, = \,\chi^{SM}_K \, \eta_K^{SM} \, \left(  \, \bar{d}_{L} s_{R} \, + \, h.c.\right) \, X
\label{eLSM}
\end{equation}
where
\begin{eqnarray}
\chi^{SM}_K & = & G_F \sqrt{2} \,\frac{3}{16 \pi^2}\, \frac{m_{s}}{\zeta_W} \, m_t^2  
\nonumber \\
\eta_K^{SM} & = & \sum_{i=t,c,u} \, \left( \frac{m_i}{m_t} \right)^2 \, V_{i d}^* \,V_{i s}
\label{etachi}
\end{eqnarray}
This is in the quark mass eigenstates basis and in the approximation 
of vanishing mass of the light quark.  A similar diagram and formula 
apply for $B \rightarrow K + \mathrm{boson}$ with the substitutions 
$s\rightarrow b$ and $d\rightarrow s$.

\begin{figure}[t]
\centering
\includegraphics[scale=0.45]{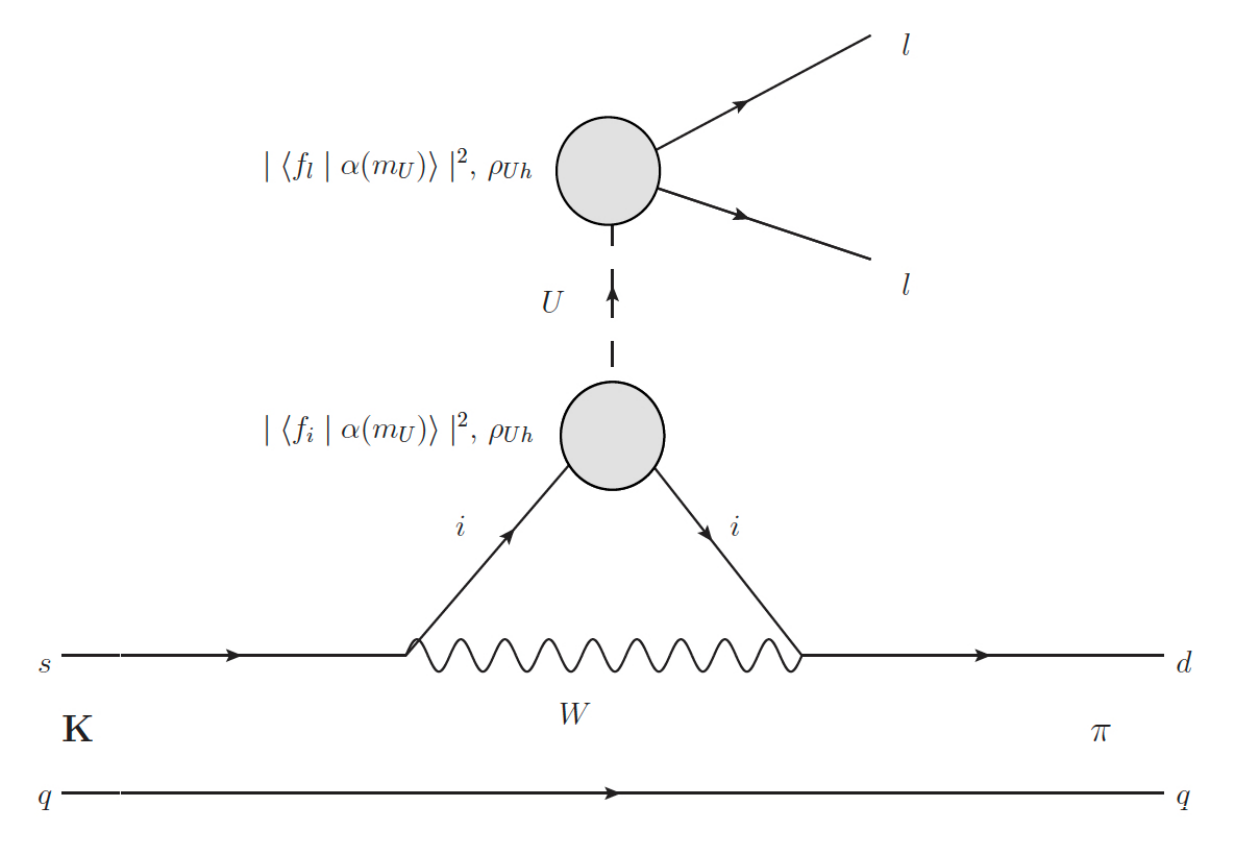}
\caption{The contribution of $U$ in FSM to the decay of the kaons, $ K \rightarrow \pi ll$, in the spectator model (\ref{Ucouplings}).}
\label{penguinU}
\end{figure}

Go over now to the FSM.  The same diagram Figure \ref{penguin-comb}
(left) will appear there, and although the formulation is different, 
the result will be the same, as shown in \cite{Higgcoup}.  But there
will be the extra contribution from $U$.

The $U$ boson, if it really exists, can contribute to the semileptonic 
decays of $K$s and $B$s via basically the same penguin diagram shown in 
Figure \ref{penguin-comb} (right).  Since the $U$ is supposed to acquire 
its couplings to quarks and leptons only through its component of size 
$\rho_{Uh}$ in the SM Higgs $h_W$, and this latter in the FSM couples 
as in (\ref{Yukawaw}), these couplings will take the form:
\begin{equation}
\mathcal{L}_Y \, = \, - \rho_{Uh} \,
\sum_{T=U,D,L,\nu} \frac{m_T}{\zeta_W} \sum_{i=1,2,3} 
  \mid\langle f_i \mid \balpha(\mu) \rangle  \mid^2\,  \overline{f_i}{f_i} \, U.
\label{Ucouplings}
\end{equation}
As it stands, $\balpha$ depends on the scale $\mu$.  For the physical 
values, we evaluate $\balpha$ at $\mu = m_U$, as is commonly thought 
appropriate, and as was done in the $g - 2$ problem in \cite{fsmanom},
giving then, for $K$ decay, the couplings indicated in Figure \ref{penguinU}.   
Hence, in place of (\ref{eLSM}) above for the SM contribution, one has 
the following effective Lagrangian from the $U$ in FSM:
\begin{equation}
\mathcal{L}_{sdU} \, = \,\rho_{Uh} \, \chi_K \, \eta_K \, \left(  \, \bar{d}_{L} s_{R} \, + \, h.c.\right) \, U
\label{effective}
\end{equation}
with $\chi_K$ unchanged but $\eta_K$ modified:
\begin{eqnarray}
\chi_K & = & \chi^{SM}_K \, = \, G_F \sqrt{2} \,\frac{3}{16 \pi^2}\, \frac{m_{s}}{\zeta_W} \, m_t^2
\nonumber \\
\eta_K  & = &  \sum_{i=t,c,u} \,  \left( \frac{m_i}{m_t} \right)^2 \, V_{i d}^*V_{i s} \,\mid \langle i \mid \balpha(m_U)\rangle \mid^2
\label{etaK}
\end{eqnarray}

A similar parametrization and result can be obtained for the $B$ 
decays with the substitutions $s\rightarrow b$ and $d\rightarrow s$. The effective Lagrangian and the parameters are:
\begin{eqnarray}
\mathcal{L}_{bsU} & = &\rho_{Uh} \, \chi_B \, \eta_B \, \left(  \, \bar{s}_{L} b_{R} \, + \, h.c.\right) \, U 
\nonumber \\
\chi_B & = & \chi^{SM}_B  \, = \, G_F \sqrt{2} \,\frac{3}{16 \pi^2}\, \frac{m_{b}}{\zeta_W} \, m_t^2
\nonumber \\
\eta_B  & = &  \sum_{i=t,c,u} \,  \frac{m_i^2}{m_t^2} \,  V_{i s}^*V_{i b} \,\mid \langle i \mid \balpha(m_U)\rangle \mid^2
\label{etaB}
\end{eqnarray}

To go from these formulae for quark decays to the actual decays of 
the mesons $K$ and $B$ will involve soft hadron physics and its 
usual uncertainties.  But these are little different from what are
already met with in the standard model contribution.  One can do 
therefore no better than follow what has been done there.  Indeed, 
since what interests us most is the comparison between the SM and 
$U$ contributions, one might hope that by treating hadron physics 
the same in both, some of the uncertainties incurred may be partly 
cancelled out.  Following then \cite{willeyyu,winkler}, we write:
\begin{equation}
\Gamma (K \rightarrow \pi U) \, = \, \mid \rho_{Uh} \mid^2 \mid \chi_K \mid^2 \, \mid \eta_K \mid^2  \, \mid \langle \pi \mid \bar{d}_R s_L \mid K \rangle \mid^2 \, 
\frac{\lambda^{1/2} (K, \pi U)}{16 \pi m_K}
\end{equation}
for kaon decays ($K^+$ or $K_L$) where the hadron matrix element and the kinematical factor are 
\begin{eqnarray}
& & \mid \langle \pi \mid \bar{d}_R s_L \mid K \rangle \mid^2  \, = \, \frac{1}{4} \frac{(m_K^2 - m_\pi^2)^2}{(m_s - m_d)^2} 
\nonumber \\
& & \lambda (K, \pi U) \, = \, \left(1-\left(\frac{m_\pi - m_U}{m_K} \right)^2 \right) \,
 \left( 1-\left(\frac{m_\pi + m_U}{m_K} \right)^2\right).
\end{eqnarray}

Similarly, following \cite{winkler,ballzwicky}, we write for B decays:
\begin{equation}
\Gamma (B \rightarrow K^{(*)} U) \, = \, \mid \rho_{Uh} \mid^2 \mid \chi_B \mid^2 \, \mid \eta_B \mid^2  \, \mid \langle K^{(*)} \mid \bar{s}_R b_L \mid B \rangle \mid^2 \, 
\frac{\lambda^{1/2} (B, K^{(*)} U)}{16 \pi m_B}.
\end{equation}
The hadron matrix elements and the kinematical factors are:
\begin{eqnarray}
& & \mid \langle K \mid \bar{s}_R b_L \mid B \rangle \mid^2  \, = \, \frac{1}{4} \frac{(m_B^2 - m_K^2)^2}{(m_b - m_s)^2} \, 
\left( \frac{0.33}{1- \frac{m_U^2}{37.5 \, Gev^2}} \right)^2
\nonumber \\
& & \mid \langle K^* \mid \bar{s}_R b_L \mid B \rangle \mid^2  \, = \, \frac{1}{4} \frac{m_B^4}{(m_b + m_s)^2} 
\left( \frac{1.36}{1- \frac{m_U^2}{27.9 \, Gev^2}} - \frac{0.99}{1- \frac{m_U^2}{36.8 \, Gev^2}}\right)^2
\nonumber \\
& & \lambda (B,K^{(*)} U) \, = \, \left(1-\left(\frac{m_{K^{(*)}} - m_U}{m_B} \right)^2 \right) \,
 \left( 1-\left(\frac{m_{K^{(*)}} + m_U}{m_B} \right)^2\right).
\end{eqnarray}

\begin{table}[ht]
\begin{spacing}{1.5}
\begin{eqnarray*}
\begin{array}{||c|c|c||}
\hline \hline
m_t= 173.07\,\mathrm{Gev} & m_b = 4.18\,\mathrm{Gev} & m_c= 1.27\,\mathrm{Gev} 
\\ \hline
m_s= 0.093 \,\mathrm{Gev} & m_{\rm light}= 0.005 \,\mathrm{Gev} & 
\\ \hline
V_{ud} = 1 - \lambda^2/2 & V_{cd} = - \lambda &  V_{td} = A \lambda^3 (1 - \rho - I \eta) 
\\ \hline
V_{us} = \lambda &  V_{cs} = 1 - \lambda^2/2 &   V_{ts} = -A \lambda^2 
\\ \hline
V_{ub} = A \lambda^3 (\rho - I \eta) & V_{cb} = A \lambda^2 &  V_{tb} = 1 
\\ \hline
A=0.814 & \lambda = 0.2257 & \rho=0.135
\\ \hline
\eta = 0.349 & &
\\ \hline
\zeta_{W}=246 \,\mathrm{Gev} & G_F = 1.166 \times 10^{-5} \,\mathrm{Gev}^{-2} & 
\\ \hline
m_K = 0.494 \,\mathrm{Gev} & m_{K^0} = 0.497\,\mathrm{Gev} & m_{K^*} = 0.892\,\mathrm{Gev} 
\\ \hline 
m_\pi = 0.140\,\mathrm{Gev} & m_{\pi^0} = 0.135\,\mathrm{Gev} & m_B=5.279\,\mathrm{Gev}
\\ \hline
\multicolumn{3}{||c||}{\Gamma_{tot}(K) = 5.36  \times 10^{-17} \,\mathrm{Gev}}
\\ \hline
\multicolumn{3}{||c||}{\Gamma_{tot}(K_L) = 1.3 \times 10^{-17} \, \mathrm{Gev}}
\\ \hline
\multicolumn{3}{||c||}{\Gamma_{tot}(B) = 4.0  \times 10^{-13} \, \mathrm{Gev}}
\\
\hline \hline
\end{array}
\end{eqnarray*} 
\caption{Input parameters used in the calculation of (\ref{BrmesonU}).}
\label{valueparameters}
\end{spacing}
\end{table}

Putting in then the numbers, one obtains the branching ratios of interest 
as follows:
\begin{eqnarray}
Br (K^+ \rightarrow \pi^+ U) \,& = & \, (1.4 \times 10^4) \, \mid \rho_{Uh}\mid^2 \, \mid\eta_K\mid^2
\nonumber \\
Br (K_L \rightarrow \pi U) \,& = & \, (6.0 \times 10^4) \, \mid\rho_{Uh}\mid^2 \, \mid Im(\eta_K) \mid^2
\nonumber \\
Br (B^+ \rightarrow K^+ U) \,& = & \,(3.0 \times 10^2)  \, \mid\rho_{Uh}\mid^2 \, \mid\eta_B\mid^2
\nonumber \\
Br (B \rightarrow K^* U) \,& = & \,(3.4 \times 10^2)  \, \mid\rho_{Uh}\mid^2 \, \mid\eta_B\mid^2.
\label{BrmesonU}
\end{eqnarray}
The numbers used in this calculation are standard but we list them in Table \ref{valueparameters} for completeness.

Next, the $U$ state being supposedly narrow---a supposition that 
will be checked in {\bf [i]} of the next section---its contribution 
can be taken just as an addition to that from the SM.  Or, in the 
language of branching ratios, the branching ratio integrated 
over all the invariant energies of the final leptonic decay products, 
is the incoherent sum of the SM contribution ($SM$) and new physics 
($NP$) consequent of the $U$ exchange.  Further, for the same reason 
that $U$ is narrow, the $NP$ piece can be split into first the decay 
of the meson emitting the $U$ scalar, followed by the decay of the 
$U$ into the lepton final state, thus, for example:
\begin{eqnarray}
Br ( K\, \rightarrow \,\pi  \,e^+\, e^-) \mid_{NP} & = & 
   Br ( K\, \rightarrow \,\pi  \,U) Br ( U\, \rightarrow  \,e^+\, e^-).
\label{completebr}
\end{eqnarray}

This gives then explicitly the expressions for the branching ratios 
for the decay modes in Table \ref{experimentaltable} for which data 
are available. Writing the difference between the experimental data and the SM results:
\begin{equation}
Br \, \mid_{exp} - Br \, \mid_{SM} \, = \, Br \mid_{NP},  
\end{equation}
we get:
\begin{eqnarray}
& & Br (K^+ \rightarrow \pi^+ \cancel{E})\mid_{NP}   =  (1.4  \times 10^{4}) \quad \rho_{Uh}^2 \mid\eta_K\mid^2  Br (U\rightarrow \cancel{E}) 
\nonumber \\
& & Br (K_L \rightarrow \pi^0 \cancel{E})\mid_{NP}  =  (6.0  \times  10^{4}) \quad \rho_{Uh}^2  \mid Im( \eta_K) \mid^2  Br (U\rightarrow \cancel{E}) 
\nonumber \\
& & Br (K_L \rightarrow \pi^0 e^+ e^-)\mid_{NP}  =  (6.0 \times 10^{4}) \quad \rho_{Uh}^2  \mid Im( \eta_K) \mid^2  Br (U\rightarrow e^+ e^-)  
\nonumber \\
& & Br (B^+ \rightarrow K^+ \cancel{E})\mid_{NP} =  (3.0 \times  10^{2}) \quad \rho_{Uh}^2  \mid\eta_B\mid^2  Br (U\rightarrow \cancel{E}) 
\nonumber \\
& & Br (B \rightarrow K^* \cancel{E})\mid_{NP}  =  (3.4 \times  10^{2}) \quad \rho_{Uh}^2  \mid\eta_B\mid^2  Br (U\rightarrow \cancel{E}) 
\nonumber \\
& & Br (B^+ \rightarrow K^+ e^+ e^-)\mid_{NP}   =  (3.0 \times  10^{2}) \quad \rho_{Uh}^2  \mid\eta_B\mid^2 \ Br (U\rightarrow e^+ e^-)
\nonumber \\
& &  Br (B \rightarrow K^* e^+ e^-)\mid_{NP}  =  (3.4 \times  10^{2}) \quad \rho_{Uh}^2  \mid\eta_B\mid^2  Br (U\rightarrow e^+ e^-) 
\nonumber \\
\label{NPBr1}
\end{eqnarray}

\section{Comparing, then combining, theory with experiment}

To make contact with experiment, one notes first that the decay 
widths of $K$s into $\pi U$ and $B$s into $K U$, as listed in 
(\ref{BrmesonU}), are calculable in the FSM once given a value 
of $m_U$ together with the information already gathered from 
earlier work.  The parameter $\rho_{Uh}$ is constrained by the 
$g - 2$ anomaly \cite{fsmanom} as seen in Figure \ref{rhoUh}, 
while the quark state vectors $|f_i \rangle$ are given in the 
Appendix of \cite{tfsm}, and $\balpha(m_U)$ for any $m_U$ can 
be read from Figure \ref{thetamu} with (\ref{curvetraced}).  
Indeed, even most of the quark masses and CKM matrix elements 
were calculated in \cite{tfsm}, but for greater precision, one 
would prefer to take those directly from experiment.  

The $U$ so produced will subsequently decay into $e^{+}  e^{-}$ 
or neutrino pairs ($U$ is below threshold for decay into $\mu^+ \mu^-$), the decay widths into which are calculable 
also in the FSM, though not so confidently for neutrino pairs 
because of uncertainties in the see-saw mechanism.
The difficulty, however, is that in the FSM $U$ can decay also 
into hidden sector particles, such as `co-neutrinos', if there 
are such with low enough masses, and widths for these modes one 
has not yet managed to estimate.  If these exist, they would be 
part of the missing energy modes $\cancel{E}$ and contribute to 
the total width of $U$.  For this reason, from earlier results 
alone, the FSM cannot as yet give the branching ratios into the 
various individual modes listed in (\ref{NPBr1}).  However, as 
we shall show later in this section, one will be able do so by 
inputting just a little bit of information from the semileptonic 
decays of $K$s and $B$s.

\begin{figure}
\begin{center}
\includegraphics[scale=0.45]{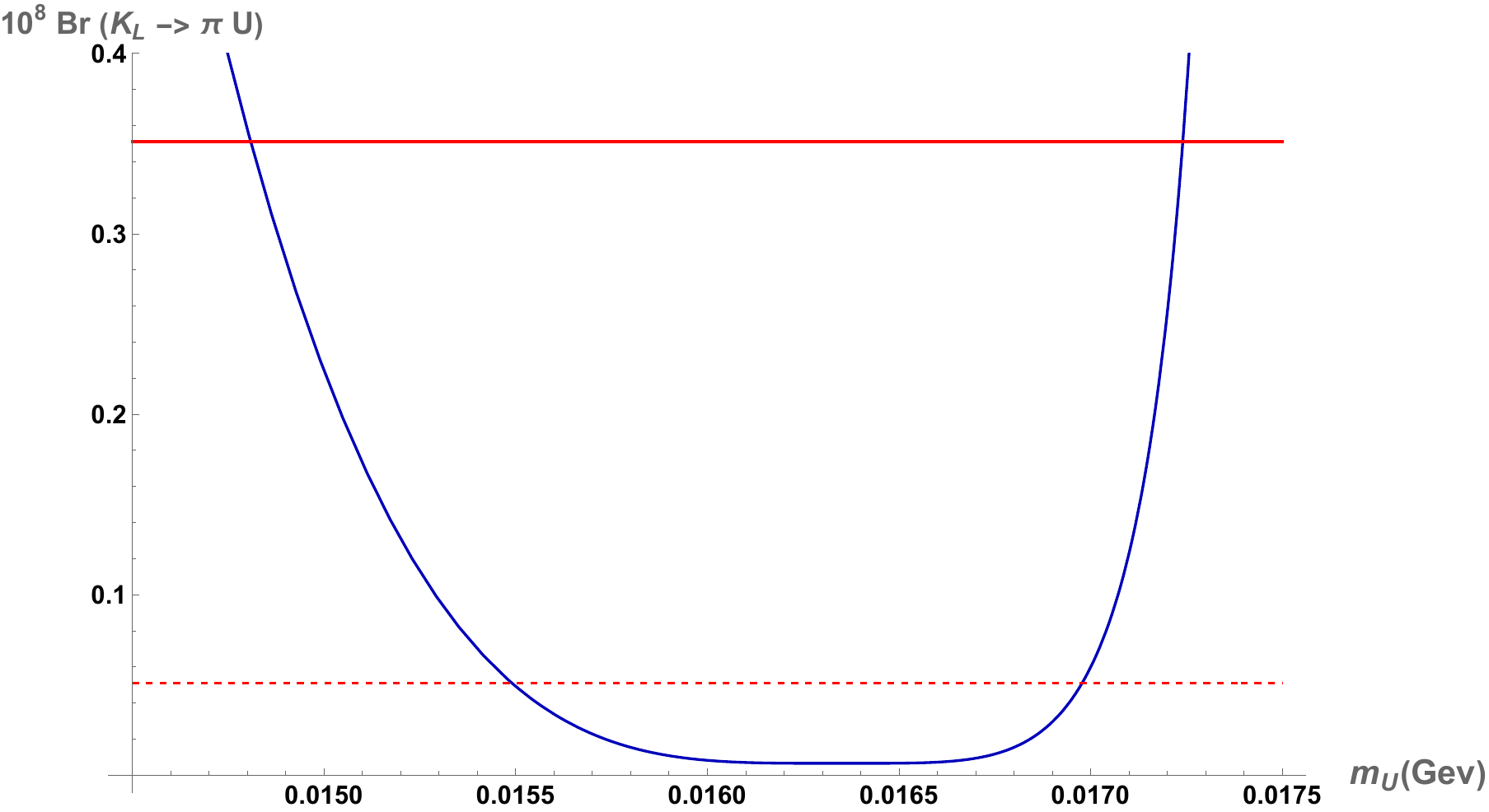}
\caption{Comparing experimental bounds on new physics to predictions of the 
$U$ proposal in FSM for the decays $K_L \rightarrow \pi U$ (blue solid line). Red solid (dashed) line: experimental bound on new physics for $K_L \rightarrow \pi$ all ($K_L \rightarrow \pi e^+ e^- $).}
\label{BrKLpiU}
\end{center}
\end{figure}

\begin{figure}
\begin{center}
\includegraphics[scale=0.45]{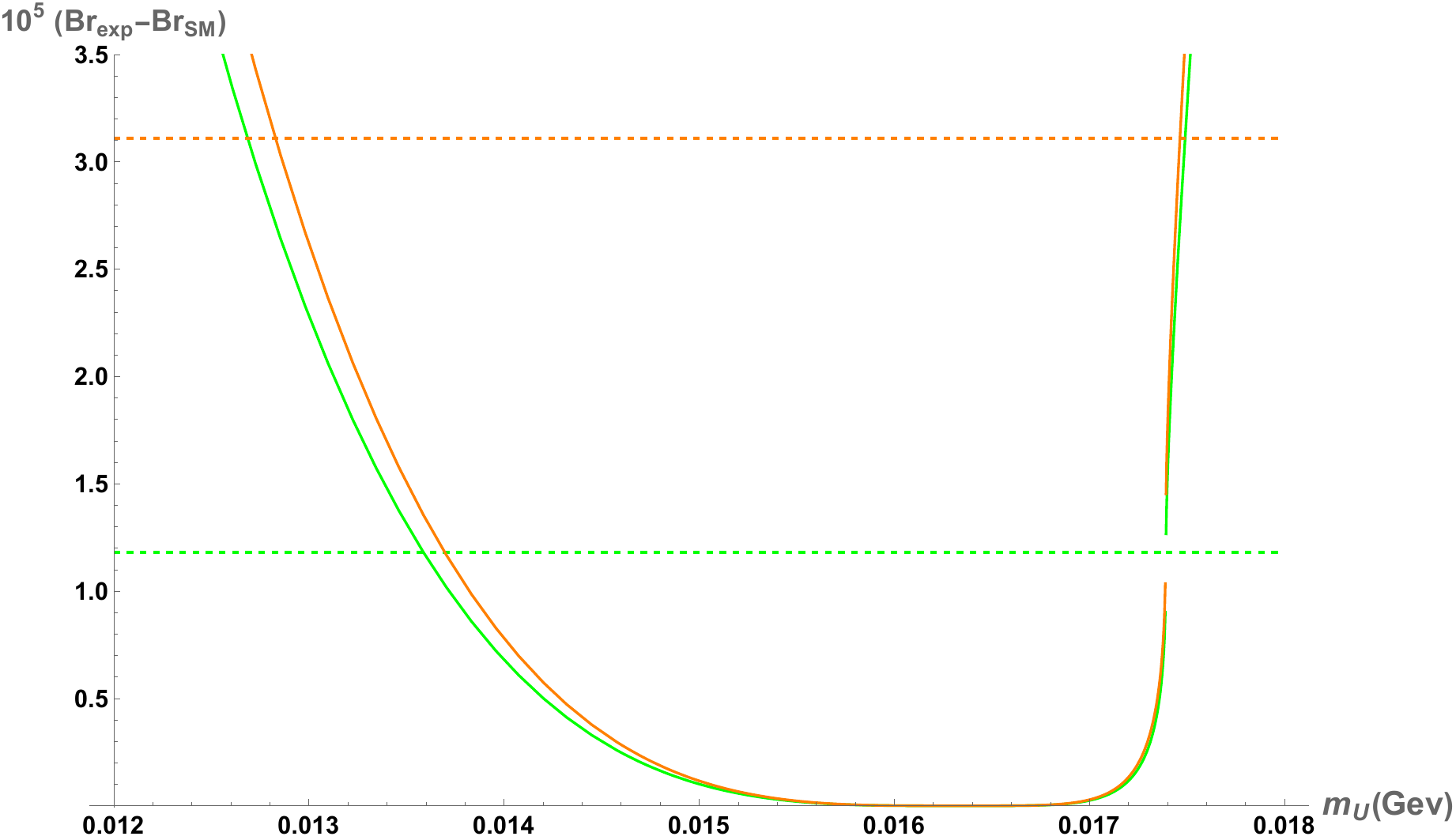}
\caption{Comparing experimental bounds (dashed lines) on new physics to predictions of the 
$U$ proposal in FSM (solid lines).  Green (Orange) for the decay $B \rightarrow K (K^*) U$.}
\label{BrBKK*U}
\end{center}
\end{figure}

Here, let us first compare the FSM result on total BRs into $U$ 
with the sums of the BRs into the $e^+ e^-$ and the $\cancel{E}$ 
modes where these are available in Table \ref{experimentaltable}, 
namely for the modes $K_L \rightarrow \pi U$, $B \rightarrow K U$ 
and $B \rightarrow K^* U$, except for the $K^+$ decay mode where 
the SM prediction for the $e^+ e^-$ mode is missing.  The result 
is shown in Figures \ref{BrKLpiU} and \ref{BrBKK*U}.

One sees that in all 3 cases the prediction of the $U$ proposal 
in FSM lies mostly outside the experimental bounds, as was expected 
with apprehension in the Preamble because of the relatively large 
value of the parameter $\rho_{Uh}$.  But, to one's surprise and big 
relief, there is an overlapping narrow region of $m_U$ just below 
17 MeV, that is, exactly where $m_U$ is thought to be, where the 
FSM prediction lie well within the experimental bounds.  Hence, we 
have the conclusions {\bf [a$'$]} and {\bf [b$'$]} already posted 
in the Preamble.  In other words, it is shown not only that the $U$ 
proposal in FSM has passed this first test posed by the semileptonic 
decays of $K$s and $B$s, but also that the data on these decays have 
supplied a tight constraint on the $U$ mass, to within a few MeV of 
the predicted unmixed value of 17 MeV.

This is, at first sight, a quite astounding result, requiring the 
remarkable coincidence that the coupling of the $U$ be very close 
to vanishing in exactly the region needed to save the FSM from 
violating existing experimental bounds.  It is clearly important 
to understand in more detail how this comes about.  One can do so 
as follows. 

\begin{figure}[ht]
\begin{center}
\includegraphics[scale=0.5]{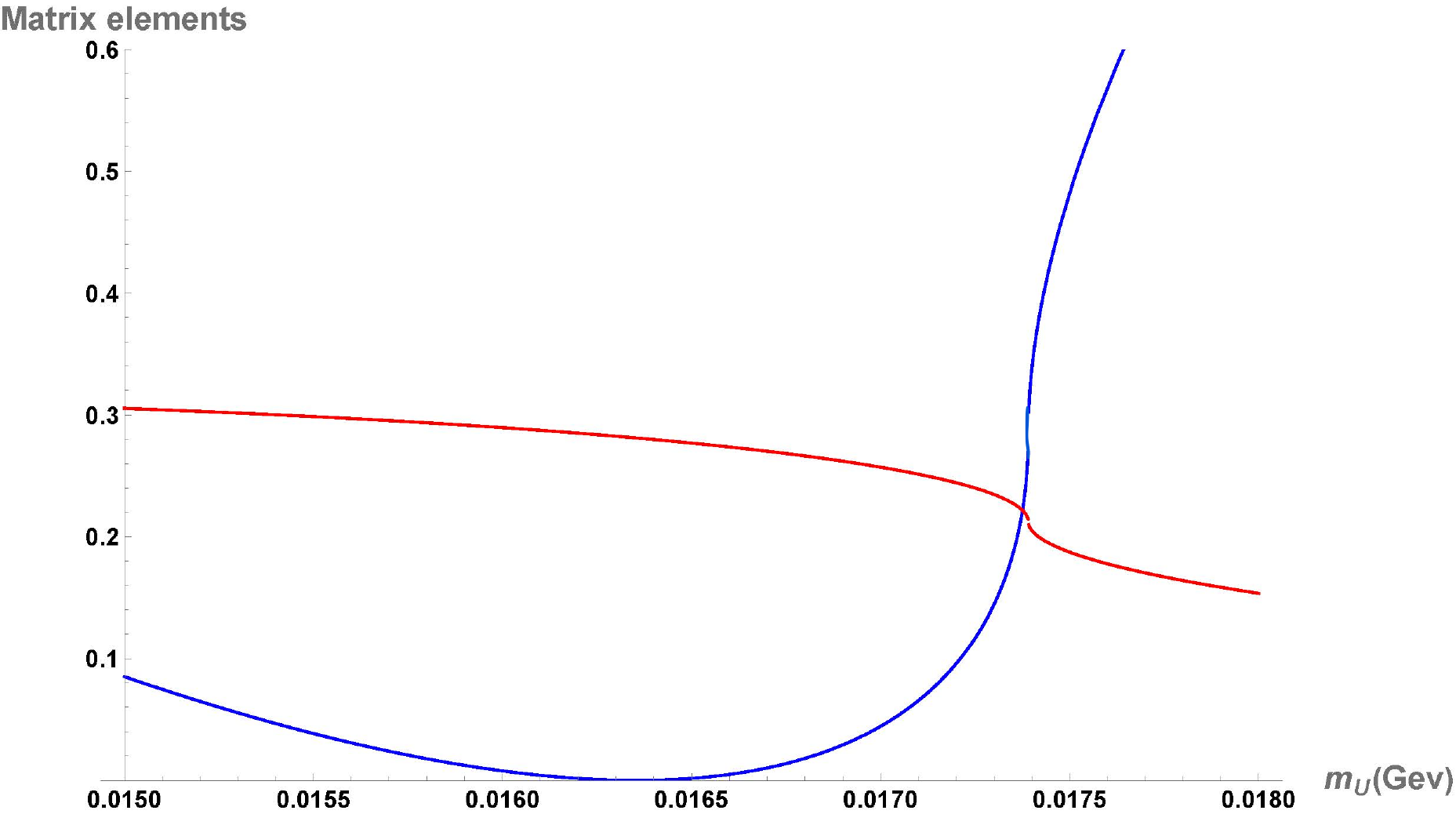}
\caption{Red: $\mid \langle \be \mid \balpha(m_U)\rangle \mid^2$. Blue: 5 $\mid \langle \bt \mid \balpha(m_U)\rangle \mid^2$}
\label{matrixelements}
\end{center}
\end{figure}

From (\ref{etaK}) and (\ref{etaB}) above, one sees that, because 
of the factor $(m_i/m_t)^2$ therein, the decay rates in all the 3 
cases shown in Figures \ref{BrKLpiU} and \ref{BrBKK*U}
are dominated by the $t$ quark in the loop.  The near vanishing 
of the whole $U$ contribution would then occur if $\balpha(m_U)$ 
happens to be near orthogonal either to the state vector $\bt$ of 
$t$ or the state vectors of the leptons to which it decays.  In 
Figure \ref{matrixelements} it is shown that for $m_U$ in the 
region of interest, $\balpha(m_U)$ is not anywhere near orthogonal 
to the state vector $\be$ of electron which is quoted as example 
for leptons, but is indeed nearly orthogonal to the state vector 
$\bt$ of $t$. 

But why should it turn out that $\balpha(m_U)$ in that narrow range 
of $m_U$ be near orthogonal to $\bt$?  To see how this happens, let 
us write $\bt$ and $\balpha(m_U)$ in polar co-ordinates as:
\begin{equation}
\bt = \balpha(m_t) =  
\left( \begin{array}{c} 
      \sin \theta_t \cos \phi_t \\
      \sin \theta_t \sin \phi_t \\
      \cos \theta_t \end{array} \right),
      \quad 
\balpha(m_U) = \left( \begin{array}{c} 
      \sin \theta_U \cos \phi_U \\
      \sin \theta_U \sin \phi_U \\
      \cos \theta_U \end{array} \right).
\label{btf}
\end{equation}

Elimination $\phi_t$ and $\phi_U$ using (\ref{curvetraced}), one 
has:
\begin{equation}
\bt = \balpha(m_t) =  
\left( \begin{array}{c} 
      - \frac{\sin \theta_t \cos \theta_t}{\sqrt{a^2 +\cos^2 \theta_t}} \\
      \frac{a \sin \theta_t }{\sqrt{a^2 +\cos^2 \theta_t}} \\
      \cos \theta_t \end{array} \right),
      \quad 
\balpha(m_U) = \left( \begin{array}{c} 
      \frac{\sin \theta_U \cos \theta_U}{\sqrt{a^2 +\cos^2 \theta_U}} \\
      - \frac{a \sin \theta_U}{\sqrt{a^2 +\cos^2 \theta_U}} \\
      \cos \theta_U \end{array} \right),
\label{btfa}
\end{equation}
account having been taken of the fact that $\phi_t$ is in the 2nd 
quadrant while $\phi_U$ is in the 4th, on the tangent plane through 
the pole looking from above.

Now, we recall that $a$ is an integration constant of a rotation 
equation for $\balpha$ governing its departure from planarity, 
for which a small value, $a = - 0,1$, was obtained in the fit to 
mass and mixing data done in \cite{tfsm}.  If we put $a = 0$, the
two vectors in (\ref{btfa}) are exactly orthogonal when $\theta_U 
= \pi/2 - \theta_t$.  Hence, since $a$ is small, one would expect 
that the two vectors to remain close to orthogonality. Indeed, inputting in (20) the fitted value
of $a = - 0.1$ and $\theta_t = 1.33$ 
rad\footnote{The difference in sign to what is read in Figure 3 can be ignored,
being due just to an unusual convention for $\theta$ adopted in \cite{tfsm} for convenience in integrating the rotation equation for $\balpha$.}, one obtains for $\theta_U = \pi/2 - \theta_t$:
\begin{equation}
\langle \bt \mid  \balpha(m_U) \rangle \sim 1.0 \times 10^{-2},
\label{nearorthog}
\end{equation}
that is $\bt$ and $\balpha(m_U)$ are very nearly orthogonal.  
Since this factor appears 
in the branching ratios in (\ref{NPBr1}) raised to the 4th power, 
it gives a suppression of order $10^{-8}$ and easily explains the 
dips seen in Figures \ref{BrKLpiU} and \ref{BrBKK*U}.  

But why should the dip be so close in scale or in $m_U$ to 17 MeV, 
the unmixed value?  This is understood when we recall that 17 MeV 
corresponds to the pole where the rotation of $\balpha$ undergoes 
the vacuum transition VTR1 where $d \theta/ d \ln \mu \sim \infty$, 
as can be seen in Figure \ref{thetamu}.  Indeed, from this figure, 
one can just about discern with good eyes that a shift of the polar 
angle, $\delta \theta = \pi/2 - \theta_t \sim 0.24$, as is required, 
will result from a shift of $\mu$ by merely 1 MeV from the pole 
position.  The same observation explains also why the constraints on 
$m_U$ obtained in Figures \ref{BrKLpiU} and \ref{BrBKK*U} are 
so sharp, since a small shift in $m_U$ is enough to displace 
$\balpha(m_U)$ from orthogonality with $\bt$, sufficiently to annul 
the suppression of the decay rates due to the factor $\mid \langle 
t \mid \balpha(m_U)\rangle \mid^2$.  

With this, we have understood in some detail and even roughly checked 
the at first sight surprising calculational results displayed in the 
Figures \ref{BrKLpiU} and \ref{BrBKK*U}.  If this is 
indeed the right interpretation of the experimental situation, it 
would have required an internal co-ordination and consistency from 
the FSM, and a degree of accuracy in the rotation trajectory of 
$\balpha$ over some 4 decades in energy from $m_t \sim 170$ GeV to 
17 MeV, to a greater extent, perhaps, than one could have expected 
from the rough quality of the fit to data done in \cite{tfsm} to 
obtain it.  It seems to point to a robustness in the formulation of 
the fit which is belied by the crudeness so far in effecting it. 

There are two points here worth noting:
\begin{itemize}
\item {\bf [1]}
The constraint {\bf [b$'$]} obtained here for $m_U$ is much sharper 
than that obtained previously \cite{fsmanom} from the Lamb shift 
anomaly, and being below the unmixed value 17 MeV, is more likely 
as a value from mixing.  However, the new constraint is in a sense 
less satisfying than a constraint from the Lamb shift anomaly if the 
latter still exists after the recent change in data.  If the Lamb 
shift anomaly exists, it will have to be explained, and if by the 
$U$, then the constraint ensues.  For the present constraint from 
semileptonic decays of $K$s and $B$s, on the other hand, there is 
no deviation as yet from the SM to be explained away.  Constraints 
arise only from the need to explain the absence of effects in those 
decays if one insists on the existence of the $U$ because of the 
$g - 2$ anomaly.  Despite the result here, therefore, one still 
awaits eagerly clarification of the experimental situation in the 
Lamb shift anomaly.
\item {\bf [2]}
The predicted value of $m_U$ in Figures \ref{BrKLpiU} and \ref{BrBKK*U}
is uncannily close to that of the anomalous signal reported, twice 
by now, by \cite{Atomki1,Atomki2} and under further phenomenological 
study \cite{X17}.
As an explanation of this Atomki 
anomaly in beryllium decay, the $U$, a $J^P = 0^+$ state, is excluded 
if parity has to be conserved, for which reason we suggested in 
\cite{fsmanom} another state for its explanation.  However, if the 
$U$ couples like the Higgs boson $h_W$, as here maintained, then 
parity may not be conserved in its coupling to standard matter, in 
which case, its possible relevance to the Atomki anomaly may have 
to be reconsidered.  
\end{itemize}

This is about all that we have been able to extract from comparing 
the data on semileptonic decays of $K$s and $B$s with the $U$ 
proposal, so long as we rely only on previous results in the FSM.  
But, if we allow ourselves to feed in some information from these 
decays themselves, then further interesting conclusions result, as 
follows.

\begin{figure}[t]
\begin{center}
\includegraphics[scale=0.45]{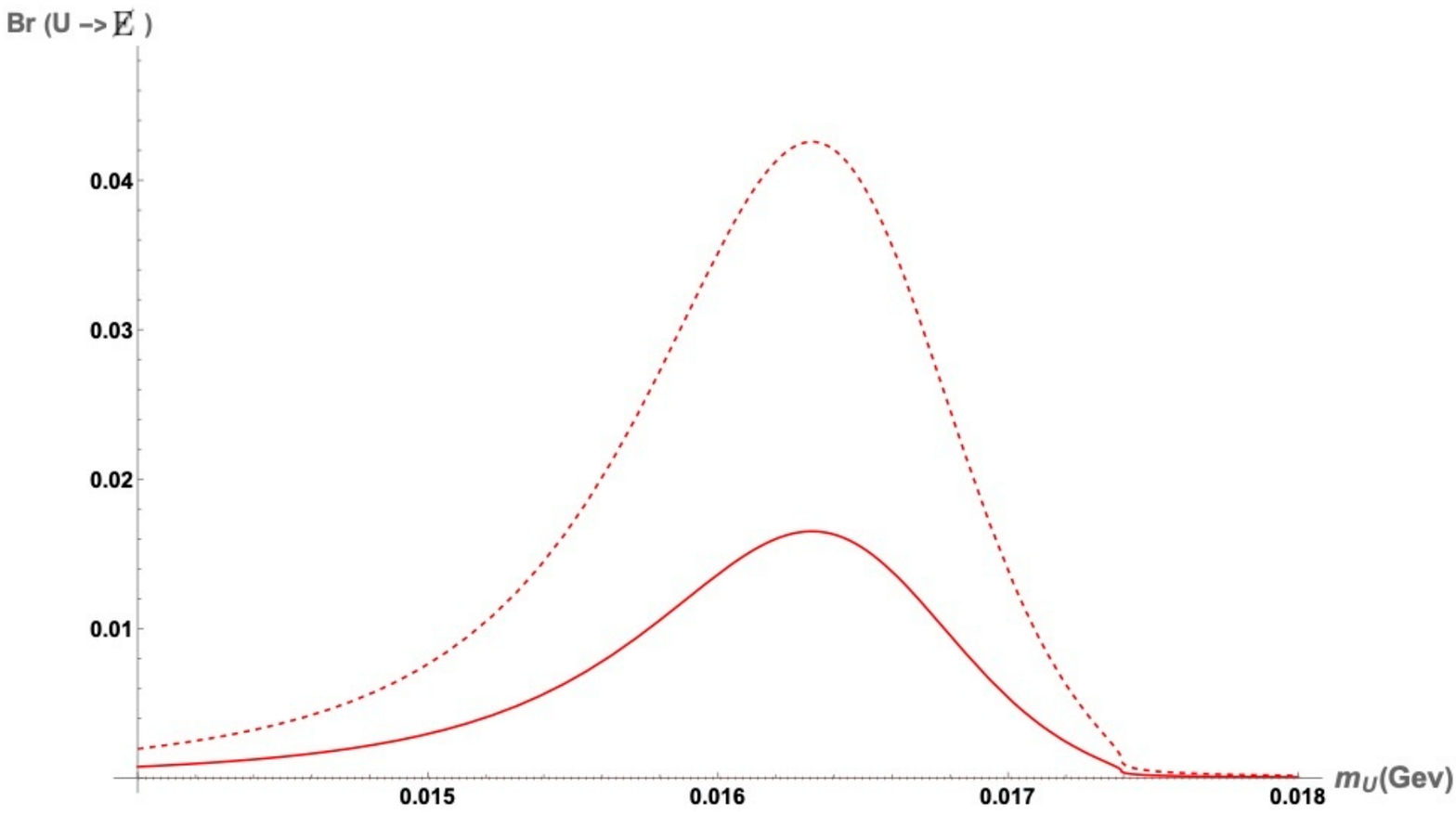}
\caption{$Br (U\rightarrow \cancel{\rm E}) $ obtained with the experimental data of the Table \ref{experimentaltable} and equation (\ref{NPBr1}). Solid line: central value. Dashed line upper limit, the lower limit being compatible
with zero.}
\label{BrUmissing}
\end{center}
\end{figure}

The so-called `golden channel' $K^+ \rightarrow  \pi^+ \, \cancel{E}$,
besides being well measured, is distinguished, among the missing energy 
channels in Table \ref{experimentaltable}, in citing an actual signal 
for new physics instead of just a bound, although the error is such as 
to admit still the SM prediction without any addition of new physics.  
If one takes these numbers seriously, and puts in (\ref{completebr})  
the already calculated FSM value for $Br ( K\, \rightarrow \,\pi  \,U)$, 
one obtains for the branching ratio $U \rightarrow \cancel{E}$ the 
estimate shown in Figure \ref{BrUmissing}, where we see that the decay 
of $U$ into neutrinos or other undetectable neutral particles has only 
a small branching ratio in the whole mass range of interest, with a 
maximum value of around 0.04 at the preferred mass of $U \sim$ 16.5 
MeV.  We note that at this value of $m_U$, the contribution of the $t$ 
quark in the loop almost vanishes, and it is here the $c$ quark in the 
loop, involving as it does some larger CKM elements, which gives the main 
contribution to the branching ratio $Br ( K\, \rightarrow \,\pi  \,U)$ 
and hence the just derived result.

This conclusion can only be regarded as tentative, depending as it does 
on just one piece of data and on some details of the FSM which have not 
been checked against experiment elsewhere.  But, if accepted, it leads 
to the following very interesting results.

First, as regards the semileptonic decays being considered and listed 
in Table \ref{experimentaltable}, it means that one can now compare 
the predictions of the $U$ proposal with experiment not just in the 
total $U$ contribution as before but separately in the 2 individual 
modes for $U$ decaying into $\cancel{E}$ and $e^+ e^-$.  For those 
modes with final missing energy the predictions are well below the 
present experimental limits, 
\begin{eqnarray}
  \nonumber \\
Br ( K_L\, \rightarrow \,\pi  \,\cancel{E}) |_{NP} &=&   Br ( K_L\, \rightarrow \,\pi  \,U) Br ( U\, \rightarrow \,\cancel{E}) \leq 4 \times 10^{-11} \ll
10^{-9}
\nonumber \\
Br ( B \, \rightarrow \,K \,\cancel{E}) |_{NP} &=&   Br ( B\, \rightarrow \,K  \,U) Br ( U\, \rightarrow \,\cancel{E}) \leq  10^{-8} \ll 10^{-5}
\nonumber \\
Br ( B\, \rightarrow \,K^*  \,\cancel{E}) |_{NP} &=&   Br ( B\, \rightarrow \,K^*  \,U) Br ( U\, \rightarrow \,\cancel{E}) \leq 10^{-8} \ll 10^{-5},
\end{eqnarray}
being some 2-3 order of magnitude below the difference between present 
experimental result and SM prediction.  Hence, a lot of increase in 
the experimental precision would be needed to extract conclusion for 
them.  

\begin{figure}
\begin{center}
\includegraphics[scale=0.45]{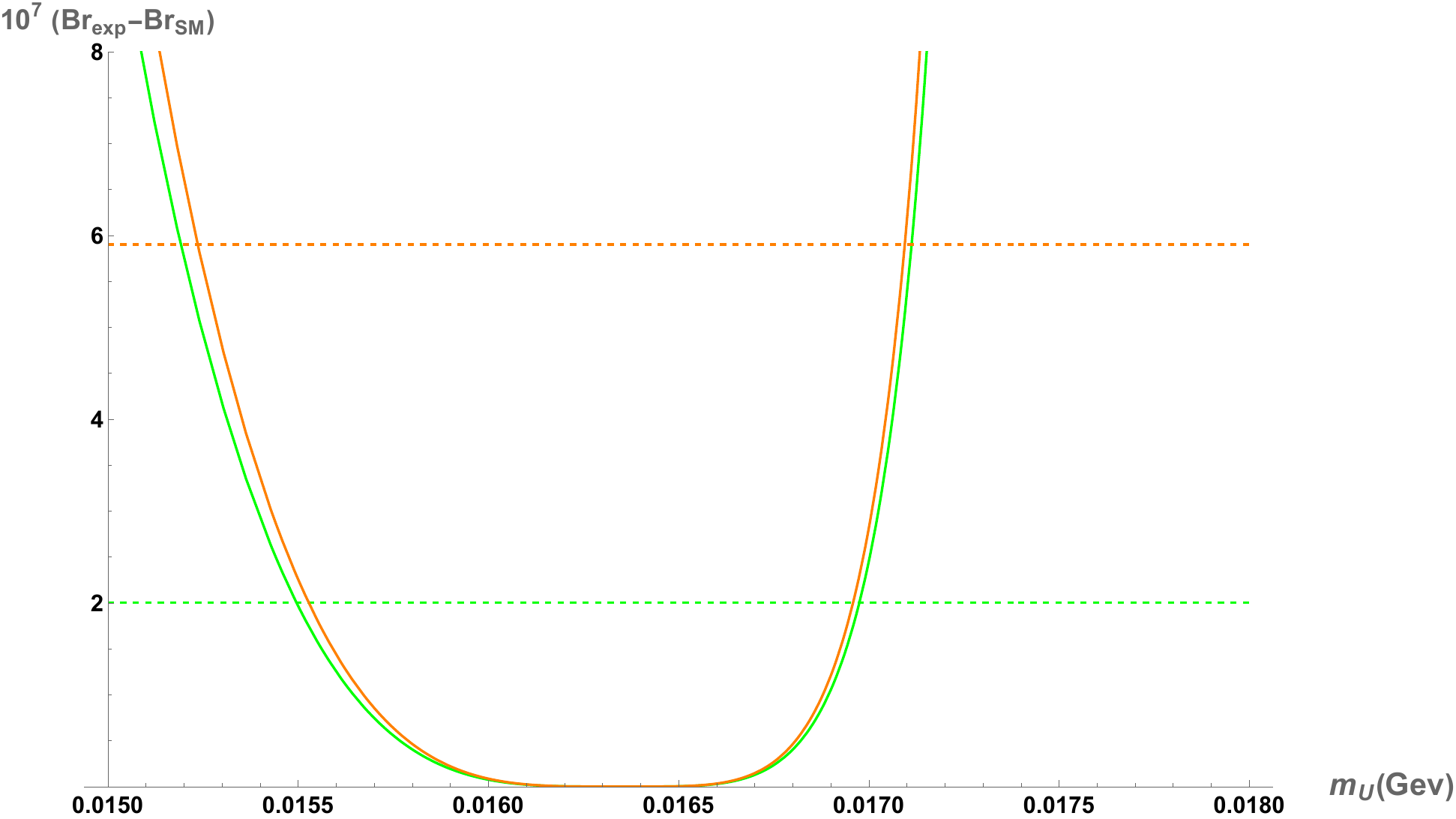}
\caption{Comparing experimental bounds (dashed lines) on new physics to predictions of the 
$U$ proposal in FSM (solid lines).Green (Orange) for the decay $B \rightarrow K (K^*) e^+ e^-$.}
\label{BrBKK*Uee}
\end{center}
\end{figure}

Further, to the extent that these small modes into missing energy can 
be ignored, the bounds on branching ratios into $U$ (total) in Figures 
\ref{BrKLpiU} and \ref{BrBKK*U} can be replaced by those where the $U$ 
decays into the $e^+ e^-$ channel only, and the implied bounds on $m_U$ 
are tightened further, as seen in Figure \ref{BrKLpiU} (dashed line) 
and Figure \ref{BrBKK*Uee}, to say 15.5 - 17.0 MeV.   

Secondly, and perhaps even more interestingly, it can lead within 
the FSM framework to the following predictions of wider potential 
significance:
\begin{itemize}
\item {\bf [i]}
Since the $U$ will now decay predominantly into the $e^+ e^-$ mode,
the known couplings in FSM of the $U$ to $e^+ e^-$ can be used 
to estimate the total width of $U$, as follows:
\begin{equation}
\Gamma (U \rightarrow e^+ e^- )\mid_{FSM}\,  = \,  
   \frac{\rho_{Uh}^2}{8 \pi} \, \mid \langle e \mid \balpha(m_U)\rangle \mid^4 \, 
   \left(\frac{m_\tau}{\zeta_W} \right)^2 
   m_U \left( 1- \left(\frac{2 m_e}{m_U}\right)^2 \right)^{3/2}
\end{equation}
or, with $\mid \langle e \mid \balpha(m_U)\rangle \mid \sim 0.45$ 
as given in \cite{tfsm} Appendix, one has
\begin{equation}
\Gamma_{Total}(U) \sim 1.7 \times 10^{-11}\ {\rm GeV},
\label{GammaUT}
\end{equation}
or a lifetime for $U$ of around $4 \times 10^{-14}$ s.

This estimate accords with the supposed narrowness of $U$ used above 
to derive (\ref{NPBr1}), and seemingly also with the limits imposed 
by experiment of the Atomki anomaly on the $U$ width in relation to 
the remark {\bf [2]} made before. 

Indeed, with its mass, width and many couplings though tentatively 
but quite accurately known, one is now likely in the position to ask 
whether the $U$ should have been seen before in experiment, and if 
not, whether one can devise some more direct means for its detection.
And if the $U$ is found, it will serve as a window into the hidden 
sector which, according to \cite{cfsm}, will be a whole new chapter, 
not just in particle physics but in cosmology as well.

\item {\bf [ii]}
The suggested bound of $<$ 4 percent on the branching ratio for $U$ 
into missing energy, though small, is still much larger than its 
branching ratio into neutrinos obtained from the FSM:
\begin{equation}
\Gamma (U \rightarrow \nu \bar{\nu})\mid_{FSM}\,  = \,  \frac{\rho_{Uh}^2}{8 \pi} \, 
   \sum_i \mid \langle \nu_i \mid \balpha(m_U)\rangle \mid^4 \, 
   \left(\frac{m_{\nu_3}}{\zeta_W} \right)^2 m_U,
\end{equation}
where $m_{\nu_3}$ is meant to be the Dirac mass of the heaviest 
neutrino, fitted in \cite{tfsm} to about 29 MeV. This gives a 
partial width into neutrinos only as:
\begin{equation}
\Gamma (U \rightarrow \nu \bar{\nu})\mid_{FSM}\,  
   \sim \, 3 \times 10^{-14}\ {\rm GeV},
\end{equation}
or a branching ratio of only 0.002. 
 
This means that some room remains, although not much, for $U$ decay 
into unknowns.  What could these be?  We recall now that $U$, as 
predicted from the FSM, is a mixed state with a small component 
proportional to $\rho_{Uh}$ in the standard sector, but a large 
component in the hidden sector.  It would prefer therefore to decay 
into hidden sector particles if there are such light enough for it 
to do so.  From the analysis of the particle spectrum performed in 
\cite{cfsm}, it seems that the most likely candidates in FSM are 
the hidden sector analogues of neutrinos in the standard sector, 
called co-neutrinos in \cite{cfsm}, only with the roles of flavour 
and colour interchanged, because these, in principle, can be subject 
to see-saw mechanisms, as neutrinos are, to end up with very small 
masses.  And if there are light co-neutrinos with mass smaller than 
half the $U$ mass, say $< 8$ MeV, then the $U$ would be expected to 
decay copiously into them, much more preferably than into $e^+ e^-$, 
suppressed as the latter is by $\rho_{Uh}^2$.  This is contrary to 
what is seen in Figure \ref{BrUmissing} with a maximum of 4 percent 
for the branching ratio $U \rightarrow \cancel{E}$.  It can happen, 
of course, that the decay of $U$ into co-neutrino pairs is itself 
suppressed by another mechanism, which is not impossible given that 
the $U$ is so near in mass to the vacuum transition VTR1 where many 
unexpected things are happening \cite{cfsm,cevt}.  Until we know of 
such, however, it seems fair to assume that if light co-neutrinos, 
say $\chi$, do exist, then the mode $U \rightarrow \chi \bar{\chi}$ 
would overwhelm the $e^+ e^-$ mode instead.  A reasonable tentative 
conclusion at this stage of our knowledge would thus be that there 
are no light co-neutrinos, or indeed any light dark matter particle 
in the hidden sector with mass less than 8 MeV.  If true, this can 
be an important fact to know for cosmology. 
\end{itemize}
It has to be said again, however, that these 2 results, though 
potentially of much significance with far-reaching consequences, 
have to be regarded as only tentative for reasons already given.

\section{Concluding remarks}

In summary:
\begin{itemize}
\item {\bf [a$'$]}
The proposed explanation \cite{fsmanom} of the $g - 2$ anomaly by a 
$U$ boson, which is a scalar state in the hidden sector predicted by 
the FSM at about 17 MeV with a small admixture of the SM Higgs $h_W$
(mixing strength $\rho_{Uh} \sim 0.1$) is shown now to be compatible 
with the existing data on semileptonic decays of $K$s and $B$s.  This 
is the point that the present paper was initially meant to check, and 
it has passed the test.
\item {\bf [b$'$]}
The result has provided as surprise bonus a quite sharp confirmation 
of the predicted mass of the $U$ state just below 17 MeV, as seen in 
Figures \ref{BrKLpiU}, \ref{BrBKK*U}, and \ref{BrBKK*Uee}.  This is an 
independent check on the FSM since the data on semileptonic decays 
of $K$s and $B$s have not, of course, even been thought of in the 
earlier work \cite{tfsm} or \cite{fsmanom}.
\item {\bf [c$'$]}
Injecting some information learnt from these semileptonic decays gives 
further that (i) $U$ has a width of order $10^{-11}$ GeV or a lifetime 
of order $10^{-14}$ s, (ii) a possible lower bound on the mass of the
lightest dark matter particle of $\sim 8$ MeV. 
\end{itemize}
In other words, the $U$ proposal in FSM seems not only to be consistent
so far both with experiment and within itself but also to have serious 
repercussions beyond.

There is, however, one special feature in the FSM, and in particular 
in the $U$ proposal, which deserves much closer theoretical scrutiny 
and is, for us, a cause for some concern.  As in the SM, masses and 
couplings of particles in the FSM acquire through renormalization a 
dependence on scale, and as in the SM also, their physical values 
are in the FSM taken to be the values when the scale equals the mass 
of the particle itself.  That is, symbolically, $m_{phys}$ of particle 
$X$ is taken to be the fixed point of the function $m_X(\mu)$, or as 
the solution to the equation:
\begin{equation}
m_X(\mu) = \mu.
\label{mphys}
\end{equation}
The difference from the standard usage in the SM lies only in that 
this criterion is applied in the FSM to renormalization by colour 
framon loops, which has no analogue in the SM.

Indeed, according to the FSM, before applying renormalization from 
framon loops, the 2 lighter generations of quarks and leptons are 
massless and there is no mixing (in itself not a bad
approximation numerically to start with), 
and it is only through the stated 
renormalization that they acquire their nonzero masses and mixings.  
In other words, the masses and mixings of the lighter generations 
in the FSM are not to be inserted, as they are in the SM, as input 
from experiment, but are to be deduced as perturbative effects from 
radiative corrections, explaining thereby their small values relative 
respectively to the heaviest generation mass and to unity.  Since 
renormalization by radiative corrections depends on scale, this 
means that the physical values of masses and mixing parameters 
would depend on the criterion (\ref{mphys}) for evaluating them. 
Now this may be a cause of some concern, for what was in the SM just 
a criterion for evaluating small corrections has been now applied 
in the FSM to define the masses and mixing angles of the 2 lighter 
generations, which erstwhile in the SM were taken as fundamental 
quantities.

Moving now, as we have done in this paper, to the $U$ proposal, this 
cause for concern gets exacerbated since $U$ is near in scale to the 
vacuum transition VTR1 at 17 MeV where the scale-dependence of the 
polar angle of $\balpha$ as seen in Figure \ref{thetamu} is infinite. 
All our conclusions are therefore strongly dependent on the criterion 
that the couplings of $U$ are to be evaluated at the scale of the $U$ 
mass.  Do we, indeed, have sufficient theoretical justification to do
so? 

Besides, there are ambiguities in applying this criterion for choosing 
scales.  Take for example the couplings of the Higgs boson $h_W$ to 
quarks and leptons.  There are 2 mass scales involved, that of $h_W$ 
and that of the fermion; then which are we to choose for evaluating 
the physical value?  In the SM, the arguments that we have seen to 
justify choosing the scale, say, for evaluating the mass of the $b$ 
quark at the mass scale of the $b$ itself, as in (\ref{mphys}) above, 
go roughly as follows.  There are, in the QCD perturbation series, 
terms in ascending powers of $\ln(\mu/m_b)$, so that, to assure the 
best convergence, one chooses $\mu = m_b$ to eliminate such terms.  
Then, having obtained the best value of $m_b$ in this way, we look 
at the coupling of $h_W$ to $b$.  This may be affected by and acquire 
scale-dependence from renormalization by electroweak loops.  To fix 
next the best value for the coupling, we then choose the scale for 
this step of the calculation to be the mass scale of the Higgs $h_W$. 
Hence, the logic would seem to be that we first choose the scale to 
guarantee the best convergence for the loops which give the strongest 
scale-dependence, in this case the QCD loops, and then, having done 
so, to do the same for the electroweak loops which give further, but 
weaker, scale-dependence.

Go over now to the FSM scenario for the same coupling.  The fermion 
mass is now affected not only by QCD loops but also by framon loops 
via $\balpha$, which give an even stronger scale-dependence than the 
QCD loops.  To ensure the best convergence then, it seems logical 
first to choose the scale at the mass scale of the fermion to define 
the fermion mass by (\ref{mphys}), and then proceed with the rest.  
This is what was done in \cite{Higgcoup}, where it was shown that we 
get the SM scenario as result, only with mass and mixing parameters, 
which the SM takes from experiment, supplanted in the FSM by the 
values obtained in the step above.

Go further now to consider the coupling of the $U$ to the quarks 
which figure in the $g - 2$ anomaly and semileptonic decays of $K$s 
and $B$s.  There are again 2 mass scales involved, that of the $U$ 
and that of the fermions.  The scale-dependences of both are now 
dominated by framons loops; then which of the two should one choose 
to evaluate it?  Now, although both come from framon loops, the dependence 
on scale of $m_U$ is much stronger than that of the fermion masses 
because $m_U$ is close to the vacuum transition VTR1 where the slope 
of $\balpha$ is infinite as seen in Figure \ref{thetamu}.  Hence, 
following the criterion advanced above, one should choose first the mass 
scale of the $U$ for the evaluation of this coupling.  And this is 
what has been done in \cite{fsmanom} for dealing with the $g - 2$ 
anomaly and above here for studying the semileptonic decays of $K$s 
and $B$s.

The argument seems so far to be consistent, but there remain 2 points,
at least, of concern.  First, one has considered so far only relatively
simple quantities like masses and couplings.  Can the same criterion 
for choosing the renormalization scale be 
pushed consistently to complicated processes of many loop of various 
types?  Secondly, even remaining at the mass and coupling level, the 
choice of scale for evaluating them is no longer a matter of small 
corrections to the masses and couplings as for the SM, but, for the 
FSM, a matter of definition for these what were erstwhile thought to 
be fundamental quantities, although they are obtained within the FSM 
framework still as perturbative effects.  Should one be satisfied with 
only such intuitive arguments as we have so far been able to supply?
We do not know.  But it seems like that the problem has always been 
there already in the SM, how to deal with the scale-dependences of 
physical quantities acquired from renormalization, only in the FSM, 
it has become more acute because of the strong scale-dependence of 
certain quantities acquired via renormalization by framon loops.

\end{document}